\documentclass[aps,apl,superscriptaddress,showpacs,reprint,floatfix]{revtex4-1}
\usepackage{graphicx}   
\usepackage{bm}     

\begin{document}

\title{Optical gain in carbon nanotubes}

\author{Etienne Gaufr\`es}
\affiliation{Institut d'Electronique Fondamentale, CNRS-UMR 8622, Universit\'e
Paris-Sud 11, 91405 Orsay, France}
\author{Nicolas Izard}
\affiliation{Institut d'Electronique Fondamentale, CNRS-UMR 8622, Universit\'e
Paris-Sud 11, 91405 Orsay, France}
\affiliation{National Institute of Advance Industrial Science and Technology
(AIST), Tsukuba, Japan}
\author{Xavier Le Roux}
\affiliation{Institut d'Electronique Fondamentale, CNRS-UMR 8622, Universit\'e
Paris-Sud 11, 91405 Orsay, France}
\author{Delphine Marris-Morini}
\affiliation{Institut d'Electronique Fondamentale, CNRS-UMR 8622, Universit\'e
Paris-Sud 11, 91405 Orsay, France}
\author{Sa\"id Kazaoui}
\affiliation{National Institute of Advance Industrial Science and Technology
(AIST), Tsukuba, Japan}
\author{Eric Cassan}
\affiliation{Institut d'Electronique Fondamentale, CNRS-UMR 8622, Universit\'e
Paris-Sud 11, 91405 Orsay, France}
\author{Laurent Vivien}
\affiliation{Institut d'Electronique Fondamentale, CNRS-UMR 8622, Universit\'e
Paris-Sud 11, 91405 Orsay, France}
\email{laurent.vivien@u-psud.fr}
\homepage{http://silicon-photonics.ief.u-psud.fr}
\date{Received \today}

\begin{abstract}
Semiconducting single wall carbon nanotubes (s-SWNT) have proved to be
promising material for nanophotonics and optoelectronics. Due to the
possibility of tuning their direct band gap and controlling excitonic
recombinations in the near-infrared wavelength range, s-SWNT can be used as
efficient light emitters. We report the first experimental demonstration of
room temperature intrinsic optical gain as high as 190~cm$^{-1}$ at a wavelength of
1.3~$\mu$m in a thin film doped with (8,7) s-SWNT. These results
constitute a significant milestone towards the development of laser sources
based on carbon nanotubes for future high performance integrated circuits.
\end{abstract}
\pacs{42.55.-f, 78.45.+h, 78.55.-m, 78.67.Ch, 81.07.De}
\maketitle
\pagebreak

Unique electronic\cite{Martel-APL, Postma-Science} and
optical\cite{Avouris-Materials} properties of Single-Wall Carbon Nanotubes
(SWNT) allow them to play a major role in futur photonic
circuits\cite{Itkis-Science, Misewich-Science, Xia-NatureNano}. Since the
discovery of photoluminescence of nanotubes encapsulated into micelle
surfactant\cite{Bachilo-Science}, research efforts in nanotube photonics have
greatly intensified, and several building blocks have already been
demonstrated, such as field-effect transistor\cite{Martel-APL} or
detectors\cite{Itkis-Science}.  Luminescence properties of carbon nanotubes
attracted a special interest, and became a leading component in various
configurations, not only as individualised entities in micelle surfactant, but
also when suspended over trenches\cite{Lefebvre-Nanolett} or deposited onto
chip surface\cite{Adam-Nanolett}. Numerous works prove that the nanotube
environment plays a primordial role in their emission
behaviour\cite{Finnie-PRL, Okazaki-Nanolett, Gaufres-OL}, and several
non‑radiative de-excitation mechanisms exist, either by Auger
recombination\cite{Wang-PRB} or energy transfers to metallic SWNT
(m-SWNT)\cite{Hertel-APA} or excitons\cite{Mortimer-PRL}. These mechanisms
lead to a strong competition between bright and dark excitons and a limited
photoluminescence lifetime\cite{Wang-PRL}. Up to now, a main challenge was to
obtain light amplification in carbon nanotubes. Here, we experimentally
demonstrate optical gain in carbon nanotubes at a wavelength of 1.3~$\mu$m at
room temperature.

Semiconducting SWNT (s-SWNT) extracted by the previously described polyfluorene
(PFO) method\cite{Izard-APL} were drop cast on glass substrate. Pure s-SWNT thin
films will be hereafter designated as sample A, while a control sample made with
a blend of s- and m-SWNT will be named sample B. Both layers have a homogeneous
carbon nanotube density and a thickness of 1~$\mu$m. The coating was made to
obtain regular and vertical edge facets\cite{Supplementary}. Absorbance spectrum
of such initial s-SWNT/PFO solution and photoluminescence spectrum of
drop-casted s-SWNT/PFO solution on glass substrate are reported in ref
\onlinecite{Gaufres-OL}. Optical properties characterized by the E$_{22}$ and
E$_{11}$ optical transitions are related to strongly bound
excitons\cite{Wang-Science}. Photons absorbed on the
E$_{22}$ optical transition create excitons at the carbon nanotube surface and
light emission is produced due to the exciton recombination through the E$_{11}$
optical transition. The polymer host matrix (PFO) presents absorption and
photoluminescence peaks in the UV-visible spectrum range, and is optically
inactive (no absorption and no emission) on the near infrared carbon nanotube absorption
and emission range, without any evidence of energy transfer towards carbon
nanotubes\cite{Nish-Nanotech}.

\begin{figure}
\includegraphics[width=8.5cm]{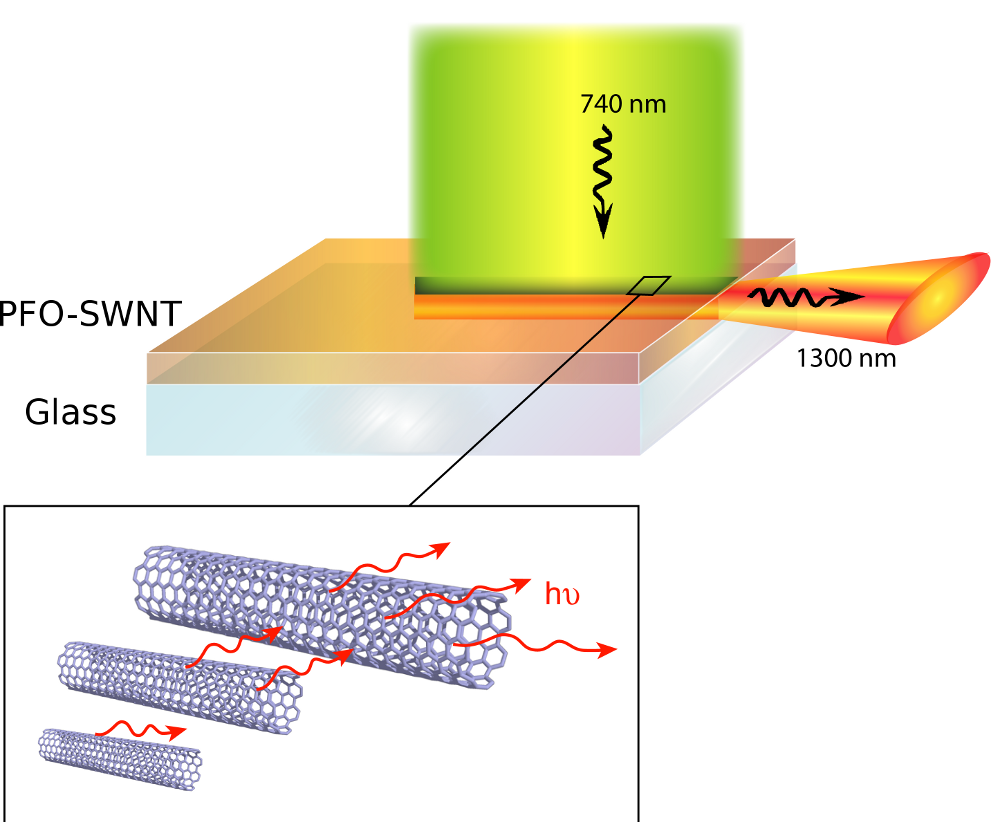}
\caption{(Color online)
Schematic representation of a Variable Scan Length (VSL) experiment.} \label{fig1}
\end{figure}
\begin{figure}
\includegraphics[width=8.5cm]{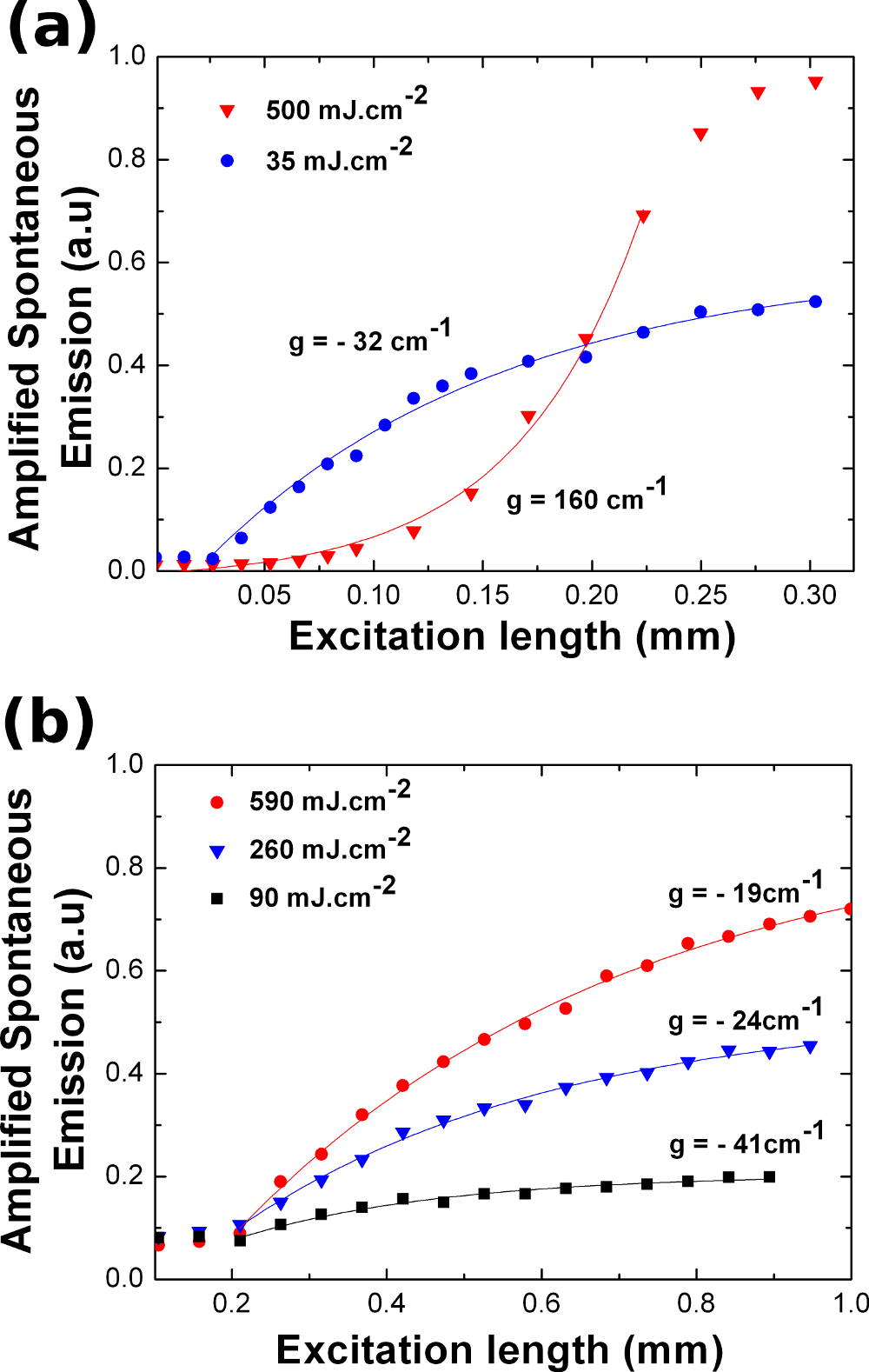}
\caption{(Color online)
\textbf{(a)} $I_{ASE}$ vs excitation
stripe length $l$ of pure s-SWNT embedded in thin layer (Sample A). Blue and
red dots correspond to two input fluences of the excitation:
35~mJ$\cdot$cm$^{-2}$ and 500~mJ$\cdot$cm$^{-2}$, respectively. Excitation and
recording wavelengths were respectively 740~nm and 1300~nm. A fit to the data
with equation \ref{eq1} is shown as a solid line. For sample A, $g_{net}$ was
140~cm$^{-1}$ in high fluence regime and -19~cm$^{-1}$ at low fluence regime.
\textbf{(b)} $I_{ASE}$ vs excitation stripe length $l$ of SWNT embedded in thin
layer composed of a mixture of s- and m-SWNT (Sample B). The input pump
fluences were explored under the same experimental condition as in
Fig.~\ref{fig2}(a).} \label{fig2}
\end{figure}

 
Optical gain measurements were first performed on sample A using the variable
strip length (VSL) method\cite{Shaklee-APL}. The sample was optically excited by
an Optical Parametric Oscillator laser emitting at a wavelength of 740~nm,
pumped by a two nanosecond pulsed Nd:YAG laser at 355~nm. Amplified Spontaneous
Emission (ASE) of carbon nanotubes was then collected at the edge of the sample
as a function of the excitation length $l$, as schematically described in
Fig.~\ref{fig1}. Fig.~\ref{fig2}(b) presents the ASE intensity ($I_{ASE}$)
of sample A as a function of length $l$, for low and high pump fluences
(35~mJ$\cdot$cm$^{-2}$ and 500~mJ$\cdot$cm$^{-2}$, respectively). At high
fluence, a strong signal increase was observed, while at low fluence, saturation
of the luminescence signal was directly reached. The difference in the ASE
responses under low and high excitations demonstrated a clear change of optical
behaviour which clearly proved optical gain feature in sample
A\cite{Shaklee-Lum, Pavesi-Nature}. $I_{ASE}$ can be described by the one
dimensional amplifier model\cite{Shaklee-Lum}:
\begin{equation}\label{eq1}
I_{ASE}(l) \sim \frac{I_{Spont}}{g_{int}-\alpha} \cdot (e^{(g_{int}-\alpha)
\cdot l} -1)
\end{equation}
were perfectly fitted as shown in Fig. 1C, where the theoretical curves for high
and low pump fluences are respectively drawn in red and blue solid lines.  The
ASE intensity evolution featured an exponential increase at high pump fluence,
while an exponential saturation at low pump fluence was obtained. The net modal
gain $g_{net}$ at high fluence was estimated to be about 160~$\pm$~10~cm$^{-1}$.
This gain value for carbon nanotubes embedded in polymer films was of the same
order of magnitude than the gain measured in other reference semiconducting
nano-material systems, such as silicon nanocrystals\cite{Pavesi-Nature}. For
long strip length excitation ($l$ $>$ 0.25~mm) and high fluence, a saturation of
$I_{ASE}$ was observed.  Main origins of this saturation were a geometrical
factor caused by the reduction of solid angle emission and the increased
absorption length by the nanotube doped layer, and a gain saturation effect
caused by the depletion of the excited state population under high ASE
intensity\cite{Negro-OptCom}. At low fluence, the intrinsic gain $g_{int}$ of
SWNT was too low to compensate for loss, and the measured net modal gain
$g_{net}$ was then negative. Using equation~\ref{eq1}, a fit to experimental
data gave an estimation of the net modal gain $g_{net}$ of about -32~cm$^{-1}$.
The same VSL experiments were also performed on sample B, which contained both
semiconducting and metallic carbon nanotube. VSL experiments were performed at
pump fluences from 90 to 590~mJ$\cdot$cm$^{-2}$, and the results are reported in
Fig.~\ref{fig2}(b). Exponential saturations were observed for each input pump
fluence which, according to equation~\ref{eq1}, corresponded to negative net
modal gain. For pump fluences from 90 to 590~mJ$\cdot$cm$^{-2}$, the net gain
varied from -41~cm$^{-1}$ to -19~cm$^{-1}$. This underlines the importance of
the absence of m-SWNT and nanoparticles in a sample on the optical gain
properties: indeed, sample A was only constituted by s-SWNT in
PFO\cite{Gaufres-OL, Izard-APL}.

\begin{figure}
\includegraphics[width=8.5cm]{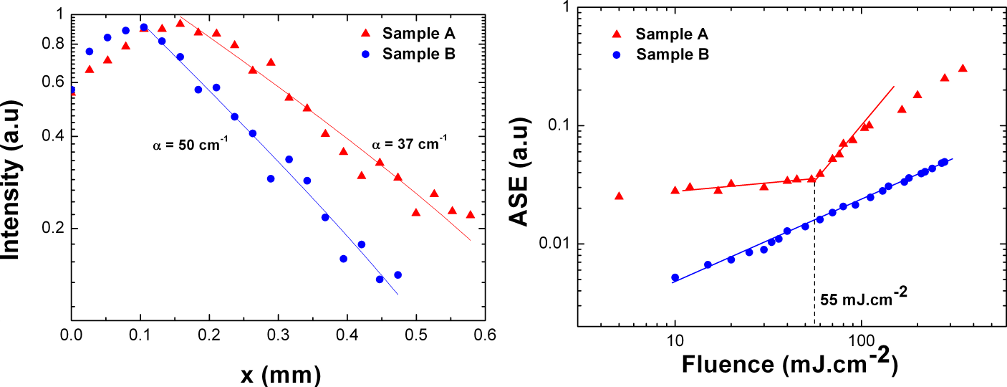}
\caption{(Color online)
\textbf{(a)} SES intensity as a function of the excitation spot position $x$ for
sample A and B. A fit with equation~\ref{eq2} gave the overall losses in the
thin layer composite. \textbf{(b)} Edge-emission intensity of the (8,7) s-SWNT at
1300~nm as a function of pump fluence at room temperature for both samples A
and B. A significant gain energy threshold was observed around
55~mJ$\cdot$cm$^{-2}$ for sample A. Solid lines are guide for the eye.}
\label{fig3}
\end{figure}

In order to provide an estimation of the overall optical losses $\alpha$, the
Shifting Excitation Spot (SES) method was used\cite{Negro-OptCom}.
Furthermore, this method allows a scan of the VSL strip length with a constant
size excitation spot and is able to demonstrate the homogeneity of the thin
SWNT based layer. SES experiments were performed on samples A and B, and
results are reported in Fig.~\ref{fig3}(a). First, $I_{ASE}$ increased and
quickly reached a maximum when the spot no more interacted with the sample
edge. After this peak, a monotone decrease was observed in log scale for both
samples, as the distance between the spot position and the sample edge
increased. First, the monotone decrease confirmed the good optical homogeneity
of both samples, and the evolution of $I_{ASE}$ is described by the following
relation\cite{Shaklee-Lum}:
\begin{equation}\label{eq2}
I_{ASE}(x) \sim I_{Spont} \cdot e^{-\alpha \cdot x}
\end{equation}
where $I_{Spont}$ is the spontaneous emission intensity per length unit and
$x$ the distance between the excitation spot and the sample edge. According to
equation~\ref{eq2} and SES measurements, loss coefficients of both samples A
and B were 37~cm$^{-1}$ and 50~cm$^{-1}$, respectively. Optical losses in
sample B were much higher than in sample A, which was directly related to the
presence of m-SWNT and nano-particles, and increasing light absorption. This
increase of loss and the less efficient light emission obtained in sample B
can explain why a positive net modal gain could not be observed. However, from
the determination of the optical loss $\alpha$ using the SES method and the
net modal gain $g_{net}$ obtained from the VSL method, the intrinsic gain
$g_{int} = g_{net} + \alpha$ for sample A was as high as
190~$\pm$10~cm$^{-1}$.

The evolution of $I_{ASE}$ as a function of the incident laser fluence was
also studied (Fig.~\ref{fig2}(b)). Experimental conditions were identical
for both samples A and B. In the case of sample A, a threshold pump fluence
was observed around 55~mJ$\cdot$cm$^{-2}$. Above the threshold, ASE intensity
presented a significant increase followed by a slight saturation. This is a
typical signature of light amplification due to an optical gain. Below the
pump fluence threshold, light emission was dominated by the losses from the
absorption peaks, while above the fluence threshold, the optical gain
dominated the losses, resulting in an enhancement of light emission. On the
contrary, such a behaviour was not observed with sample B for which the ASE
intensity behaviour was only driven by the optical loss.

\begin{figure}
\includegraphics[width=8.5cm]{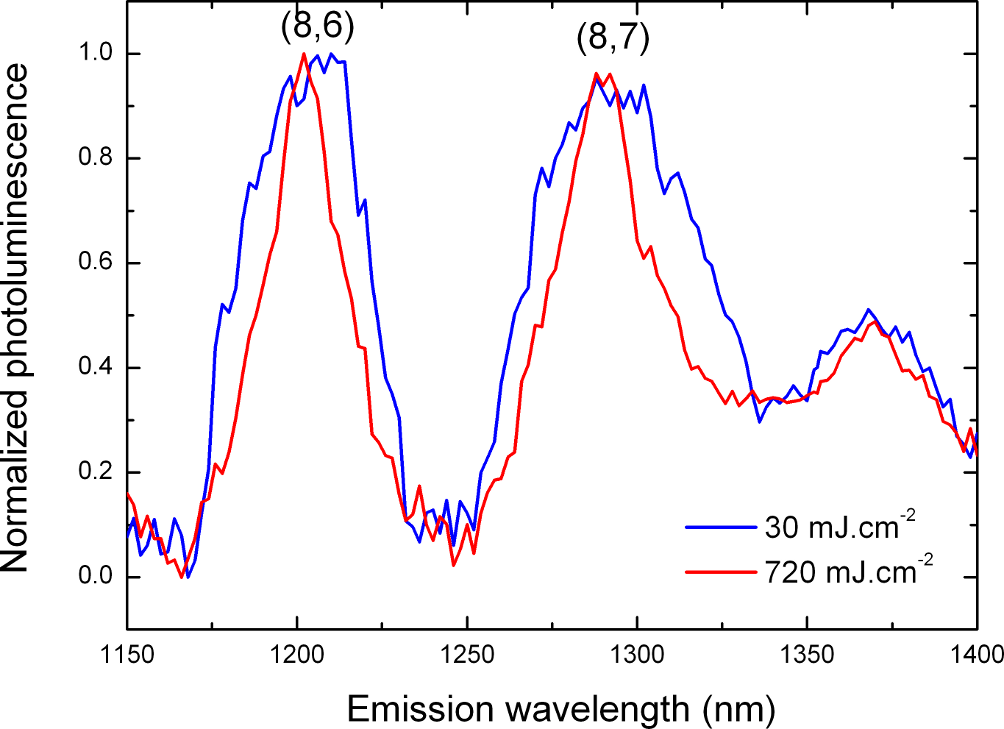}
\caption{(Color online)
High resolution normalized spectra of the ASE for sample A at low
(30~mJ$\cdot$cm$^{-2}$) and high (720~mJ$\cdot$cm$^{-2}$) pump fluences with
an excitation wavelength of 740~nm. A 29~\% linewidth narrowing (FWHM from 63
to 45~nm) was observed on the photoluminescence of (8,7) nanotube at 1300~nm,
and linewidth narrowing of 28~\% (FWHM from 44 to 32~nm) was also observed for
the (8,6) nanotube at 1200~nm. Both spectra intensity were normalized to 1,
considering the signal around 1170~nm as zero\cite{Supplementary}}.
\label{fig4}
 \end{figure}

Normalized photoluminescence spectra of nanotube in sample A for two input
pump fluences are reported in Fig.~\ref{fig4}. Both fluences were chosen to be
below and above the fluence threshold as determined in Fig.~\ref{fig3}(b).
A linewidth narrowing of 29~\% was observed for the (8,7) nanotube at 1300~nm.
This result also constituted a clear demonstration of optical gain in sample
A\cite{Quist-APL}. No linewidth narrowing was observed for sample B. We could
also observe a similar linewidth narrowing feature in (8,6) nanotube at
1200~nm, providing evidence for optical gain at several wavelengths.

 These experimental results clearly indicate significant optical gain in pure
semiconducting carbon nanotubes embedded in host polymer thin film. A few
metallic carbon nanotubes mixed with s-SWNT can strongly affect the emission
behavior and, as observed in sample B, prohibit the detection of optical gain.
The ability to generate and detect and intrinsic gain of 190~cm$^{-1}$ in
carbon nanotubes is the first step toward the development of carbon
nanotubes-based laser and more generally toward photonic circuits based on
carbon nanotubes.

\acknowledgments We thank R. Colombelli and S. Laval for
insights and fruitful discussion, A. Charrier for her help with graphic arts,
D. Riehl from DGA for his help with beam profile measurements and C.T. Cheung
for her thoroughly proofreading of the manuscript. N. Izard thanks the Japan
Society for Promotion of Science and CNRS for financial support.


\end{document}